\documentclass{PoS}
\usepackage[T1]{fontenc} % if needed
\usepackage{dcolumn}
\usepackage{bm}
\usepackage{slashed}
\usepackage{amsmath,graphicx}
\usepackage{float}
\usepackage{nicefrac}
\usepackage[normalem]{ulem}
\usepackage{subfigure} 
\usepackage{bbold} 
\usepackage[makeroom]{cancel}
\def\be{\begin{equation}}
\def\ee{\end{equation}}
\newcommand{\bel}[1]{\begin{eqnarray}\label{#1}}
\newcommand{\eel}{\end{eqnarray}}
\def\barr{\begin{array}}
	\def\earr{\end{array}}
\def\beq{\begin{eqnarray}}
\def\eeq{\end{eqnarray}}
\def\bfig{\begin{figure}}
	\def\efig{\end{figure}}

\newcommand{\nn}{\nonumber}
\newcommand{\f}[2]{\frac{#1}{#2}}
\newcommand{\onehalf}{{\nicefrac{1}{2}}}

\newcommand{\p}{\partial}

\newcommand{\tr}{{\rm tr}}
\newcommand{\rf}[1]{Eq.~(\ref{#1})}

\newcommand{\rfn}[1]{(\ref{#1})}

\def\fplusrsxp{f^+_{rs}(x,p)}

\def\fminusrsxp{f^-_{rs}(x,p)}

\def\SmunuU{{\Sigma}^{\mu\nu}}

\def\S0iU{{\Sigma}^{0i}} 
 
\def\SmnU{{\Sigma}^{\mu\nu}}

\def\ubarrp{{\bar u}_r(p)}

\def\usp{u_s(p)}

\def\urp{u_r(p)}

\def\vbarsp{{\bar v}_s(p)}

\def\vrp{v_r(p)}
\def\bmu{\beta_\mu}

\def\n0{n_{(0)}}
\def\e0{\varepsilon_{(0)}}
\def\P0{P_{(0)}}

\def\omnL{\omega_{\mu\nu}}
\def\omnU{\omega^{\mu\nu}}

\def\omnUD{\tilde {\omega}^{\mu\nu}}

\def\pmu{p^\mu}

\def\pv{{\boldsymbol p}}

\def\Wxk{{\cal W}(x,k)}
\def\Weqxk{{\cal W}_{\rm eq}(x,k)}
\def\Weqpxk{{\cal W}^{+}_{\rm eq}(x,k)}
\def\Weqmxk{{\cal W}^{-}_{\rm eq}(x,k)}
\def\Weqpmxk{{\cal W}^{\pm}_{\rm eq}(x,k)}
\def\Fxk{{\cal F}(x,k)}

\def\Feqpmxk{{\cal F}^{\pm}_{\rm eq}(x,k)}
\def\Pxk{{\cal P}(x,k)}

\def\Peqpmxk{{\cal P}^{\pm}_{\rm eq}(x,k)}
\title{Wigner function approach to polarization-vorticity coupling and hydrodynamics with spin}

\ShortTitle{Wigner function approach to polarization-vorticity ...}

\author{\speaker{Avdhesh Kumar}
\\
        Institute of Nuclear Physics Polish Academy of Sciences, PL-31342 Krakow, Poland\\
        E-mail: \email{avdhesh.kumar@ifj.edu.pl}}

\abstract{Newly introduced equilibrium Wigner functions for
	 particles with spin one-half are used in the semi-classical kinetic equations to study a possible relation between thermal vorticity and spin polarization. It is shown that in global equilibrium both the thermal-vorticity and spin polarization tensors are constant but not necessarily equal. In the case of local equilibrium, we define a procedure leading to hydrodynamic equations with spin. We introduce such equations for the  de~Groot, van~Leeuwen, and van~Weert (GLW) formalism as well as for the canonical scheme (these two frameworks differ by the definitions of the energy-momentum and spin tensors). It is found that the GLW and  canonical versions are connected by a pseudo-gauge transformation.
	}

\FullConference{XIII Quark Confinement and the Hadron Spectrum - Confinement2018\\
		31 July - 6 August 2018\\
		Maynooth University, Ireland}

\begin{document}

%%%%%%%%%%%%%%%%%%%%%%%%%%%%%%%%%%%%%%%%%%%%
\section{Introduction}
\label{intro}
Recent measurements of the $\Lambda$--hyperon  spin polarization in heavy-ion collisions by the STAR experiment~\cite{STAR:2017ckg,Adam:2018ivw} have inspired a broad interest in the theoretical studies related to spin polarization and vorticity formation. The results of various investigations that refer to:  the spin-orbit coupling~\cite{Betz:2007kg,Liang:2004ph,Liang:2004xn,Gao:2007bc,Chen:2008wh}, statistical properties of matter in global equilibrium \cite{Zubarev:1979,Becattini:2009wh,Becattini:2012tc,Becattini:2013fla,Becattini:2015nva,Hayata:2015lga}, kinetic models of spin dynamics~\cite{Gao:2012ix,Chen:2012ca,Fang:2016vpj,Fang:2016uds}, hydrodynamics with triangle anomalies~\cite{Son:2009tf,Kharzeev:2010gr} and the Lagrangian formulation of hydrodynamics  ~\cite{Montenegro:2017rbu,Montenegro:2017lvf,Montenegro:2018bcf} have been reported in this context. 

          A natural framework that can deal simultaneously with polarization and vorticity (dubbed below the hydrodynamics with spin) was proposed in Refs.~\cite{Florkowski:2017ruc,Florkowski:2017dyn}, see also Refs.~\cite{Florkowski:2018myy, Florkowski:2018fap}. This framework is based on a generalized form of the equilibrium distribution functions for particles with spin-$\onehalf$ (scalar phase-space distribution functions are replaced by 2$\times$2 relativistic spin density matrices).

         In this contribution we discuss the results of our recent work  \cite{Florkowski:2018ahw}.  We introduce the equilibrium Wigner functions for particles with spin-$\onehalf$, which satisfy the semi-classical kinetic equation, and study a possible relation between spin polarization and thermal vorticity. We also discuss a  procedure leading to the hydrodynamics with spin, for the case of the de~Groot, van~Leeuwen, and van~Weert formalism (GLW) and the canonical formalism. 
%
%
%%%%%%%%%%%%%%%%%%%%%%%%%%%%%%%%%%%%%%%%%%%%
\section{Basic concepts ---global and local thermodynamic equilibrium}
\label{sec:equilibrium}
In the case of spinless particles, the phase space distribution function $f(x,p)$ satisfies the Boltzmann equation 
\beq
p^\mu \p_\mu f(x,p) = C[f(x,p)],
\label{simpkeq}
\eeq
where $p^\mu = (E_p, \pv)$ and $\p_\mu=(\p_t,\nabla)$ are the particle four momentum and space-time derivative, while $C[f]$ is the collision integral.   The latter vanishes in the case of free streaming particles as well as in global or local thermodynamic equilibrium. In the free-streaming case, the distribution function $f_{\rm fs}(x,p) $ exactly satisfies the drift equation $p^\mu \p_\mu f_{\rm fs}(x,p) = 0$.  In the global thermodynamic equilibrium, the drift equation is also satisfied. In this case it leads to the constraints on the hydrodynamic parameters which specify the form of the equilibrium distribution $f_{\rm eq}(x,p)$. In particular, the parameters $\xi(x)$ (defined as the ratio of the local chemical potential $\mu(x)$  to the local temperature  $T(x)$) and $\beta_\mu(x)$ (defined as the ratio of the local four fluid velocity $u_\mu(x)$  to the local temperature $T(x)$ ) satisfy the conditions: $\p_\mu \xi=0$ and $\p_\mu \beta_\nu(x) + \p_\nu \beta_\mu(x) = 0$. The first equation implies that $\xi(x) = \mu(x)/T(x) = \xi^0 = \hbox{const}$. The second one is known as the Killing equation. It has a solution of the form 
\beq
\beta_\mu(x) =  \beta^0_\mu + \varpi^0_{\mu \nu} x^\nu,
\label{Killingsol}
\eeq
where, $\beta^0_\mu$ and the antisymmetric tensor $\varpi^0_{\mu \nu}$ are constants. For a given form of the $\beta_\mu(x)$ field, thermal vorticity is defined as
\bel{thvor}
\varpi_{\mu \nu} = -\frac{1}{2} \left(\p_\mu \beta_\nu - \p_\nu \beta_\mu \right).
\eel
Using \rf{Killingsol} in Eqs.~\rfn{thvor}, one can show that $\varpi_{\mu \nu}=\varpi^0_{\mu \nu}$, {\it i.e.},  the thermal vorticity in the global equilibrium is constant.

In the local equilibrium, the drift equation does not vanish. This is so, because in this case a correction $\delta f$ has to be added to the equilibrium function $f_{\rm eq}$ to describe dissipative phenomena. However, if the gradients of the local hydrodynamic variables are small, the dissipative effects can be neglected. Hydrodynamic parameters may be constrained in this case by some  specific moments of Eq.~\rfn{simpkeq} in momentum space.  They yield the conservation laws for charge, energy, and momentum. 

For particle with spin, one makes use of the Wigner functions $\Weqpmxk$ which, in addition to the standard hydrodynamic parameters, depend on an antisymmetric spin polarization tensor $\omega_{\mu\nu}(x)$. This allows us to distinguish between four rather than two different types of equilibrium: 1) {\it{global equilibrium---}} in this case the $\beta_\mu$ field is a Killing vector, $\varpi_{\mu \nu} = -\frac{1}{2} \left(\p_\mu \beta_\nu - \p_\nu \beta_\mu \right) = \omega_{\mu\nu}=\hbox{const}$, $\xi = \hbox{const}$,  2) {\it{ extended global equilibrium ---}} $\beta_\mu$ is a Killing vector, $\varpi_{\mu \nu}= \hbox{const}$, $\omega_{\mu\nu}= \hbox{const}$ but $\omega_{\mu\nu} \neq \varpi_{\mu \nu}$, $\xi = \hbox{const}$, 3) {\it{ local equilibrium ---}} $\beta_\mu$ field is not a Killing vector but one can still have  $\omega_{\mu\nu}(x) = \varpi_{\mu \nu}(x)$, $\xi = \xi(x)$, 4) {\it{ extended local equilibrium ---}} $\beta_\mu$ field is not a Killing vector, $\omega_{\mu\nu}(x) \neq  \varpi_{\mu \nu}(x)$, $\xi = \xi(x)$.

We emphasize that similarly to the case of spinless particles, the global and extended global equilibrium states correspond to the case where $\Weqxk$ exactly satisfies the kinetic equation with a vanishing collision term, 
 while for the local and extended local equilibrium states only certain moments of the kinetic equation for $\Weqxk$ can be set equal to zero (again with a vanishing collision term),  which results in the conservations laws for energy, linear and angular momentum, and charge. 

%%%%%%%%%%%%%%%%%%%%%%%%%%%%%%%%%%%%%%%%%%%%
\section{Equilibrium Wigner functions}
\label{sec:eqwignerfunctions}
In our approach we make use of the semi-classical connection between Wigner functions and the phase-space dependent  spin density matrices $f^{\pm}_{rs}(x,p)$, introduced by de Groot, van Leeuwen, and van Weert in Ref.~\cite{deGroot:1980} as follows
\beq
\Weqpxk &=& \frac{1}{2} \sum_{r,s=1}^2 \int dP\,
\delta^{(4)}(k-p) u^r(p) {\bar u}^s(p) f^+_{rs}(x,p),\nn \\
\Weqmxk &=&-\frac{1}{2} \sum_{r,s=1}^2 \int dP\,
\delta^{(4)}(k+p) v^s(p) {\bar v}^r(p) f^-_{rs}(x,p).\nn
\eeq	
Here $dP = \frac{d^3p}{(2 \pi )^3 E_p}$ is the Lorentz invariant measure in momentum space with $E_p = \sqrt{m^2 + \pv^2}$ being the on-mass-shell particle energy. Note that four momentum $k=k^\mu = (k^0, \mathbf{k})$ appearing as an argument of the Wigner function is not necessarily on the mass shell.

The equilibrium Wigner functions can be constructed by taking, as an input, the following expressions for $\fplusrsxp$ and $\fminusrsxp$ ~\cite{Becattini:2013fla}
\beq
\fplusrsxp =
\frac{1}{2m} \ubarrp X^+ \usp, ~~~~~~~~ \fminusrsxp=- \frac{1}{2m}\vbarsp X^- \vrp.  \nn
\eeq	
Here $m$ is the (anti-)particle mass, $\urp, $ and $\vrp$ are Dirac bispinors with spin indices $r$ and $s$ running from 1 to 2. The matrices $X^{\pm}$ are defined by the formula $X^{\pm} =  \exp\left[\pm \xi(x) - \bmu(x) \pmu \right] M^\pm$ where  $M^\pm=\exp\left[\pm \f{1}{2} \omnL(x)  \SmunuU \right]$ and $\SmunuU = (i/4) [\gamma^\mu,\gamma^\nu]$ is known as the Dirac spin operator. 

As it was shown in Ref.~\cite{Florkowski:2017dyn}, if we assume that the spin polarization tensor $\omnL$ satisfies the conditions, $\omnL \omnU \geq 0$ and $\omnL \omnUD = 0$, where $\omnUD=\frac{1}{2}\epsilon_{\mu\nu\alpha\beta}\omega^{\alpha\beta}$ is the dual spin polarization tensor, we can derive the following expression for the matrix $M^\pm$, 
\bel{eq:Mpmexp}
M^\pm &=& \cosh(\zeta) \pm  \f{\sinh(\zeta)}{2\zeta}  \, \omnL \SmunuU, \quad \hbox{where}\quad \zeta  = \f{1}{2} \sqrt{ \frac{1}{2} \omnL \omnU }. \nn
\eel
The parameter $\zeta$ can be interpreted as the ratio of spin chemical potential $\Omega$ to the temperature $T$~\cite{Florkowski:2017ruc}.

Wigner functions are 4$\times$4 matrices which satisfy the relation $\Weqpmxk =\gamma_0 \Weqpmxk^\dagger \gamma_0$. Therefore, they can always be decomposed in terms of the $16$ independent generators of the Clifford algebra 
\begin{equation}
\Weqpmxk=\f{1}{4} \left[ \Feqpmxk + i \gamma_5 \Peqpmxk + \gamma^\mu {\cal V}^\pm_{{\rm eq}, \mu}(x,k) + \gamma_5 \gamma^\mu {\cal A}^\pm_{{\rm eq}, \mu}(x,k)
+ \SmnU {\cal S}^\pm_{{\rm eq},\mu \nu}(x,k) \right]. \label{eq:equiwfn}
\end{equation}
Various coefficient functions appearing in the expansion of equilibrium Wigner function can be obtained by contracting $\Weqpmxk$ with appropriate gamma matrices and then taking the trace, for details see Ref.~\cite{Florkowski:2018ahw}. 
The total Wigner function is $\Weqxk = \Weqpxk + \Weqmxk. \label{totalwigner} $
%
%%%%%%%%%%%%%%%%%%%%%%%%%%%%%%%%%%%%%%%%%%%%
\section{Semi-classical expansion and Boltzmann-like equations for particles with spin} 
\label{sec:kinetic}
For an arbitrary Wigner function, a similar decomposition can be done in terms of the expansion coefficients  $\Fxk$, $\Pxk$, ${\cal V}_{\mu}(x,k)$, ${\cal A}_{\mu}(x,k)$ ,and ${\cal S}_{\mu \nu}(x,k)$.  The function $\Wxk$ satisfies the equation of the form \cite{Vasak:1987um,Florkowski:1995ei}
\bel{eq:eqforW}
\left(\gamma_\mu K^\mu - m \right) {\cal W}(x,k) = 0,  \quad\quad     {K^\mu = k^\mu + \frac{i \hbar}{2} \,\p^\mu.}\label{eq:eqforW}
\eel
The above equation exactly holds in global equilibrium and should give the constraint on hydrodynamic variables $\mu$, $T$, $u^{\mu}$ and $\omega_{\mu\nu}$ in local equilibrium. Solution of above equation can be written in the form of a series in $\hbar$,
\beq
{\cal X} = {\cal X}^{(0)}  + \hbar {\cal X}^{(1)}  +  \hbar^2 {\cal X}^{(2)}   + \cdots. \quad 	\nn \qquad \qquad {\cal X} \in \{{\cal F}, {\cal P}, {\cal V}_\mu,{\cal A}_\mu,  {\cal S}_{\nu\mu} \}
\eeq
Keeping the terms up to the  first order in $\hbar$, we can obtain the following equations for the coefficients functions ${\cal F}_{(0)}(x,k)$ and ${\cal A}^\nu_{(0)} (x,k)$,
\bel{eq:kineqs}
k^\mu \p_\mu {\cal F}_{(0)}(x,k) = 0,\quad k^\mu \p_\mu \, {\cal A}^\nu_{(0)} (x,k) = 0, 
\quad k_\nu \,{\cal A}^\nu_{(0)} (x,k) = 0.\nn
\eel
It can be easily shown that the functions ${\cal F}^{(0)}$ and ${\cal A}^{(0)}_\mu$ are basic independent ones and other coefficient functions can be expressed in terms of these two functions. 
%Therefore, we do not need to write the kinetic equations for other coefficient functions.
 Also, the algebraic structure of the equilibrium coefficient functions is consistent with the zeroth-order equations obtained from the semi-classical expansion of the Wigner function. Therefore, one can take ${\cal X}^{(0)}={\cal X}_{\rm eq}$. In this way one can get,
\bel{eq:kineqFC1}
k^\mu \p_\mu {\cal F}_{\rm eq}(x,k) = 0, \quad k^\mu \p_\mu \, {\cal A}^\nu_{\rm eq} (x,k) = 0,
\quad k_\nu \,{\cal A}^\nu_{\rm eq}(x,k) = 0.
\eel
Using the expressions (obtained by contracting $\Weqpmxk$ with appropriate gamma matrices and then taking the trace), for the coefficients functions ${\cal F}_{\rm eq}(x,k)$ and ${\cal A}^\nu_{\rm eq} (x,k)$ and then substitutes them into  Eqs.~\rfn{eq:kineqFC1}, one can see that the resulting equations will be exactly fulfilled if $\beta^{\mu}$ field satisfies the Killing equation. This suggests that thermal vorticity $\varpi_{\mu \nu}$ is constant, while the parameters $\xi$ and spin polarization tensor $\omega_{\mu \nu}$ are also constant. Note that no conclusion can be drawn if $\omega_{\mu \nu}$ is equal to $\varpi_{\mu \nu}$. This situation corresponds to case of extended global equilibrium.

%%%%%%%%%%%%%%%%%%%%%%%%%%%%%%%%%%%%%%%%%%%%
\section{Procedure to formulate hydrodynamics with spin} 
\label{sec:formulation}
\subsection{Charge current}
The charge current ${\cal N}^\alpha (x)$ can be expressed in terms of the Wigner function $\Wxk$  as follows~\cite{deGroot:1980}
\beq
{\cal N}^\alpha (x) 
&=&  \tr \int d^4k \, \gamma^\alpha \, \Wxk
=   \int d^4k \, {\cal V}^\alpha (x,k).
\label{eq:Nalphacal1}
\eeq
Using the expression for ${\cal V}^\alpha(x,k)$ up to the first order in $\hbar$, one can find the equilibrium charge current to be of the form ${\cal N}^\alpha_{\rm eq} (x) = N^\alpha_{\rm eq}(x) + \delta  N^\alpha_{\rm eq}(x)$, where $\delta  N^\alpha_{\rm eq}(x)$ is the first order in $\hbar$ correction to ${\cal N}^\alpha_{\rm eq} (x)$. It can be easily shown that $ \p_\alpha \, \delta N^\alpha_{\rm eq}(x)  = 0$. Thus, the conservation law of charge can be expressed by the equation $\p_\alpha N^\alpha_{\rm eq}(x)  = 0$, where  $N^\alpha_{\rm eq}(x)=\int d^4k \, {\cal V}^\alpha_{(0)} (x,k)$. It turns out, that this expression matches with that derived in Ref.~\cite{Florkowski:2017ruc}.

\subsection{Energy-mometum and spin tensors}

In the GLW formulation, the energy-momentum and spin tensors  are expressed in terms of the Wigner function as ~\cite{deGroot:1980}
\beq
T^{\mu\nu}_{\rm GLW}(x)&=&\frac{1}{m}\tr \int d^4k \, k^{\mu }\,k^{\nu }\Wxk=\frac{1}{m} \int d^4k \, k^{\mu }\,k^{\nu } {\cal F}(x,k), \label{eq:tmunu1}\\
S^{\lambda , \mu \nu }_{\rm GLW}(x) &=&\frac{\hbar}{4} \, \int d^4k \, \tr \left[ \left( \left\{\sigma ^{\mu \nu },\gamma ^{\lambda }\right\}+\frac{2 i}{m}\left(\gamma ^{[\mu }k^{\nu ]}\gamma ^{\lambda }-\gamma ^{\lambda }\gamma ^{[\mu }k^{\nu ]}\right) \right) \Wxk \right].\label{eq:Smunulambda_de_Groot1}
\eeq
%\bel{eq:tmunu1}
%T^{\mu\nu}_{\rm GLW}(x)=\frac{1}{m}\tr \int d^4k \, k^{\mu }\,k^{\nu }\Wxk=\frac{1}{m} \int d^4k \, k^{\mu }\,k^{\nu } {\cal F}(x,k),
%%
%\eel
%%
%\bel{eq:Smunulambda_de_Groot1}
%S^{\lambda , \mu \nu }_{\rm GLW} =\frac{\hbar}{4} \, \int d^4k \, \tr \left[ \left( \left\{\sigma ^{\mu \nu },\gamma ^{\lambda }\right\}+\frac{2 i}{m}\left(\gamma ^{[\mu }k^{\nu ]}\gamma ^{\lambda }-\gamma ^{\lambda }\gamma ^{[\mu }k^{\nu ]}\right) \right) \Wxk \right].
%\eel
%
Keeping terms only up to the first order in $\hbar$, replacing  ${\cal F}_{(0)}(x,k)$ by ${\cal F}_{\rm eq}(x,k)$ = 0, and carrying out the momentum integrations, we can reproduce the perfect-fluid formula for the  GLW energy-momentum tensor reported in Ref.~\cite{Florkowski:2017ruc}. It should obey the conservation law $\p_{\mu}T^{\mu\nu}_{\rm GLW}(x)=0$. The expression for the GLW spin tenor can be obtained by replacing $ \Wxk=\Weqxk$ and carrying out the momentum integration. Note, that if the energy-momentum tensor $T^{\mu\nu}_{\rm GLW}(x)$ is symmetric, the conservations of the orbital and spin parts of the total angular momentum should hold separately, therefore, we must have $\p_{\lambda}S^{\lambda , \mu \nu }_{\rm GLW}=0$. 

The canonical versions of the energy-momentum $T^{\mu\nu}_{\rm can}(x)$ and spin $S^{\lambda , \mu \nu }_{\rm can}(x)$ tensors can be obtained from the Dirac Lagrangian by applying the Noether theorem~\cite{Itzykson:1980rh} and are given by the following expressions
\bel{eq:tmunu1can1}
T^{\mu\nu}_{\rm can}(x)= \int d^4k \,k^{\nu } {\cal V}^\mu(x,k),
\eel
\beq
S^{\lambda , \mu \nu }_{\rm can}(x) &=& \frac{\hbar}{4} \, \int d^4k \,\text{tr}\left[ \left\{\sigma ^{\mu \nu },\gamma ^{\lambda }\right\} \Wxk  \right] 
=\frac{\hbar}{2} \epsilon^{\kappa \lambda \mu \nu} \int d^4k \, {\cal A}_{ \kappa}(x,k) \equiv \frac{\hbar}{2} \epsilon^{\kappa \lambda \mu \nu} \, {\cal A}_{ \kappa}(x).
\label{eq:Smunulambda_canonical1}
\eeq
In this case, using \rf{eq:tmunu1can1}, we obtain $T^{\mu\nu}_{\rm eq, can}(x) = T^{\mu\nu}_{\rm eq, GLW}(x) + \delta T^{\mu\nu}_{\rm can}(x)$, where $\delta T^{\mu\nu}_{\rm eq, can}(x)  = -\frac{\hbar}{2m} \int d^4k k^\nu \partial_\lambda {\cal S}^{\lambda \mu}_{\rm eq}(x,k) = -\partial_\lambda S^{\nu , \lambda \mu }_{\rm GLW}(x).$
Note that canonical energy-momentum tensor should be conserved as well, {\it i.e.}, we must have $\p_\alpha T^{\alpha\beta}_{\rm eq, can}(x) = 0$. Since $S^{\nu , \lambda \mu }_{\rm GLW}(x)$ is antisymmetric in the indices $\lambda$ and $\mu$, therefore, $\partial_\mu \, \delta T^{\mu\nu}_{\rm can}(x) = 0$. Thus,  the conservation law for the canonical energy-momentum tensor is analogous to the GLW case.

The canonical version of equilibrium spin tensor can be obtained by considering the axial-vector component in \rf{eq:Smunulambda_canonical1} in the zeroth order (with the assumption that ${\cal A}^{(0)}_{ \kappa}(x,k)= {\cal A}_{\rm eq,  \kappa}(x,k)$) and then carrying out the integration over the four-momentum $k$. It can be shown that 
\beq
S^{\lambda , \mu \nu }_{\rm can}  
&=& S^{\lambda , \mu \nu }_{\rm GLW} + S^{\mu , \nu \lambda }_{\rm GLW}+ S^{\nu , \lambda \mu }_{\rm GLW}
\label{eq:Smunulambda_canonical2}
\eeq
and 
\beq
\p_\lambda S^{\lambda , \mu \nu }_{\rm can}(x) 
= -\partial_\lambda S^{\mu , \lambda \nu }_{\rm GLW}(x) + \partial_\lambda S^{\nu , \lambda \mu }_{\rm GLW}(x)= T^{\nu\mu}_{\rm can} - T^{\mu\nu}_{\rm can}. 
\label{eq:Scancon}
\eeq
This is interesting, as one can see that the divergence of the canonical spin tensor is equal to the difference of the energy-momentum components (provided the GLW spin tensor is conserved). This result is expected because the energy-momentum tensor is not symmetric in the canonical case.

It is important to note that the two approaches (GLW and canonical) are connected via a pseudo-gauge transformation. In fact, if we define a super-potential $\Phi^{\lambda, \mu\nu} \equiv S^{\mu , \lambda \nu }_{\rm GLW}-S^{\nu , \lambda \mu }_{\rm GLW}$,
%
%\bel{Phi}
%\Phi^{\lambda, \mu\nu} \equiv S^{\mu , \lambda \nu }_{\rm GLW}-S^{\nu , \lambda \mu }_{\rm GLW},
%\eel
%
we can show that
\bel{psg1}
S^{\lambda , \mu \nu }_{\rm can}= S^{\lambda , \mu \nu }_{\rm GLW} -\Phi^{\lambda, \mu\nu}   \quad\hbox{and}\quad T^{\mu\nu}_{\rm can} = T^{\mu\nu}_{\rm GLW} + \frac{1}{2} \partial_{\lambda}\left(
\Phi^{\lambda, \mu\nu}+\Phi^{\mu, \nu \lambda} + \Phi^{\nu, \mu \lambda} \right).
\eel
%
%and
%%
%\bel{psg2}
%T^{\mu\nu}_{\rm can} = T^{\mu\nu}_{\rm GLW} + \frac{1}{2} \partial_{\lambda}\left(
%\Phi^{\lambda, \mu\nu}+\Phi^{\mu, \nu \lambda} + \Phi^{\nu, \mu \lambda} \right).
%\eel

\subsubsection{Conservation laws from the kinetic equations}
Conservation laws for charge and energy-momentum can be obtained respectively by taking the zeroth and first moments of the kinetic equation $k^\mu \p_\mu {\cal F}_{\rm eq}(x,k) = 0$. However, the equations for charge and energy-momentum are not closed due additional degrees of freedom arising from the spin polarization. To close them, we need to determine the dynamics of spin. If we multiply the kinetic equation  $k^\alpha \p_\alpha \, {\cal A}^\mu_{\rm eq} (x,k) = 0$, by a factor $\epsilon^{\mu\beta\gamma\delta}k_{\beta}$ and then integrate over $k$ we can obtain the dynamics of spin. It agrees, in fact, with the conservation of the GLW spin tensor.
\section{Summary and conclusions} \label{sec:summary}
Using the equilibrium distribution functions of spin-$\onehalf$ particles that have been put forward in Ref.\cite{Becattini:2013fla} we have constructed the equilibrium Wigner functions that satisfy the semi-classical kinetic equation. For a collision-less case, using the semi-classical expansion of Wigner function we obtain the Boltzmann like kinetic equations with spin. Using these kinetic equations we have shown that there is no direct relation between the thermal vorticity and spin polarization, except for the fact that the two should be constant in global equilibrium.  Finally, we outline the procedure to construct the conservation laws for hydrodynamics with spin within the framework of de~Groot, van~Leeuwen, and van~Weert (GLW) and in the canonical framework. In the GLW case, the  energy-momentum tensor is symmetric and spin is conserved, while for the canonical case the energy-momentum tensor is asymmetric and spin is not conserved. Interestingly, the two cases are found to be connected by the pseudo-gauge transformation.  
\section*{Acknowledgements}
AK thanks Wojciech Florkowski and Radoslaw Ryblewski for very fruitful collaboration. This research was supported in part by the Polish National Science Center Grant No. 2016/23/B/ST2/00717.

\bibliography{pv_ref}{}

\providecommand{\href}[2]{#2}\begingroup\raggedright\begin{thebibliography}{10}

\bibitem{STAR:2017ckg}
{\bf STAR} Collaboration, L.~Adamczyk et~al., {\it {Global $\Lambda$ hyperon
  polarization in nuclear collisions: evidence for the most vortical fluid}},
  {\em Nature} {\bf 548} (2017) 62--65,
  [\href{http://arxiv.org/abs/1701.06657}{{\tt arXiv:1701.06657}}].

\bibitem{Adam:2018ivw}
{\bf STAR} Collaboration, J.~Adam et~al., {\it {Global polarization of
  $\Lambda$ hyperons in Au+Au collisions at $\sqrt{s_{_{NN}}}$ = 200 GeV}},
  \href{http://arxiv.org/abs/1805.04400}{{\tt arXiv:1805.04400}}.

\bibitem{Betz:2007kg}
B.~Betz, M.~Gyulassy, and G.~Torrieri, {\it {Polarization probes of vorticity
  in heavy ion collisions}},  {\em Phys. Rev.} {\bf C76} (2007) 044901,
  [\href{http://arxiv.org/abs/0708.0035}{{\tt arXiv:0708.0035}}].

\bibitem{Liang:2004ph}
Z.-T. Liang and X.-N. Wang, {\it {Globally polarized quark-gluon plasma in
  non-central A+A collisions}},  {\em Phys. Rev. Lett.} {\bf 94} (2005) 102301,
  [\href{http://arxiv.org/abs/nucl-th/0410079}{{\tt nucl-th/0410079}}].
  [Erratum: Phys. Rev. Lett.96,039901(2006)].

\bibitem{Liang:2004xn}
Z.-T. Liang and X.-N. Wang, {\it {Spin alignment of vector mesons in
  non-central A+A collisions}},  {\em Phys. Lett.} {\bf B629} (2005) 20--26,
  [\href{http://arxiv.org/abs/nucl-th/0411101}{{\tt nucl-th/0411101}}].

\bibitem{Gao:2007bc}
J.-H. Gao, S.-W. Chen, W.-T. Deng, Z.-T. Liang, Q.~Wang, and X.-N. Wang, {\it
  {Global quark polarization in non-central A+A collisions}},  {\em Phys. Rev.}
  {\bf C77} (2008) 044902, [\href{http://arxiv.org/abs/0710.2943}{{\tt
  arXiv:0710.2943}}].

\bibitem{Chen:2008wh}
S.-W. Chen, J.~Deng, J.-H. Gao, and Q.~Wang, {\it {A General derivation of
  differential cross-section in quark-quark scatterings at fixed impact
  parameter}},  {\em Front. Phys. China} {\bf 4} (2009) 509--516,
  [\href{http://arxiv.org/abs/0801.2296}{{\tt arXiv:0801.2296}}].

\bibitem{Zubarev:1979}
D.~Zubarev, A.~Prozorkevich, and S.~Smolyanskii, {\it {Derivation of nonlinear
  generalized equations of quantum relativistic hydrodynamics}},  {\em Teor.
  Mat. Fiz.} {\bf 40} (1979) 394.

\bibitem{Becattini:2009wh}
F.~Becattini and L.~Tinti, {\it {The Ideal relativistic rotating gas as a
  perfect fluid with spin}},  {\em Annals Phys.} {\bf 325} (2010) 1566--1594,
  [\href{http://arxiv.org/abs/0911.0864}{{\tt arXiv:0911.0864}}].

\bibitem{Becattini:2012tc}
F.~Becattini, {\it {Covariant statistical mechanics and the stress-energy
  tensor}},  {\em Phys. Rev. Lett.} {\bf 108} (2012) 244502,
  [\href{http://arxiv.org/abs/1201.5278}{{\tt arXiv:1201.5278}}].

\bibitem{Becattini:2013fla}
F.~Becattini, V.~Chandra, L.~Del~Zanna, and E.~Grossi, {\it {Relativistic
  distribution function for particles with spin at local thermodynamical
  equilibrium}},  {\em Annals Phys.} {\bf 338} (2013) 32--49,
  [\href{http://arxiv.org/abs/1303.3431}{{\tt arXiv:1303.3431}}].

\bibitem{Becattini:2015nva}
F.~Becattini and E.~Grossi, {\it {Quantum corrections to the stress-energy
  tensor in thermodynamic equilibrium with acceleration}},  {\em Phys. Rev.}
  {\bf D92} (2015) 045037, [\href{http://arxiv.org/abs/1505.07760}{{\tt
  arXiv:1505.07760}}].

\bibitem{Hayata:2015lga}
T.~Hayata, Y.~Hidaka, T.~Noumi, and M.~Hongo, {\it {Relativistic hydrodynamics
  from quantum field theory on the basis of the generalized Gibbs ensemble
  method}},  {\em Phys. Rev.} {\bf D92} (2015), no.~6 065008,
  [\href{http://arxiv.org/abs/1503.04535}{{\tt arXiv:1503.04535}}].

\bibitem{Gao:2012ix}
J.-H. Gao, Z.-T. Liang, S.~Pu, Q.~Wang, and X.-N. Wang, {\it {Chiral Anomaly
  and Local Polarization Effect from Quantum Kinetic Approach}},  {\em Phys.
  Rev. Lett.} {\bf 109} (2012) 232301,
  [\href{http://arxiv.org/abs/1203.0725}{{\tt arXiv:1203.0725}}].

\bibitem{Chen:2012ca}
J.-W. Chen, S.~Pu, Q.~Wang, and X.-N. Wang, {\it {Berry Curvature and
  Four-Dimensional Monopoles in the Relativistic Chiral Kinetic Equation}},
  {\em Phys. Rev. Lett.} {\bf 110} (2013), no.~26 262301,
  [\href{http://arxiv.org/abs/1210.8312}{{\tt arXiv:1210.8312}}].

\bibitem{Fang:2016vpj}
R.-H. Fang, L.-G. Pang, Q.~Wang, and X.-N. Wang, {\it {Polarization of massive
  fermions in a vortical fluid}},  {\em Phys. Rev.} {\bf C94} (2016), no.~2
  024904, [\href{http://arxiv.org/abs/1604.04036}{{\tt arXiv:1604.04036}}].

\bibitem{Fang:2016uds}
R.-H. Fang, J.-Y. Pang, Q.~Wang, and X.-N. Wang, {\it {Pseudoscalar
  condensation induced by chiral anomaly and vorticity for massive fermions}},
  {\em Phys. Rev.} {\bf D95} (2017), no.~1 014032,
  [\href{http://arxiv.org/abs/1611.04670}{{\tt arXiv:1611.04670}}].

\bibitem{Son:2009tf}
D.~T. Son and P.~Surowka, {\it {Hydrodynamics with Triangle Anomalies}},  {\em
  Phys. Rev. Lett.} {\bf 103} (2009) 191601,
  [\href{http://arxiv.org/abs/0906.5044}{{\tt arXiv:0906.5044}}].

\bibitem{Kharzeev:2010gr}
D.~E. Kharzeev and D.~T. Son, {\it {Testing the chiral magnetic and chiral
  vortical effects in heavy ion collisions}},  {\em Phys. Rev. Lett.} {\bf 106}
  (2011) 062301, [\href{http://arxiv.org/abs/1010.0038}{{\tt
  arXiv:1010.0038}}].

\bibitem{Montenegro:2017rbu}
D.~Montenegro, L.~Tinti, and G.~Torrieri, {\it {The ideal relativistic fluid
  limit for a medium with polarization}},  {\em Phys. Rev.} {\bf D96} (2017),
  no.~5 056012, [\href{http://arxiv.org/abs/1701.08263}{{\tt
  arXiv:1701.08263}}].

\bibitem{Montenegro:2017lvf}
D.~Montenegro, L.~Tinti, and G.~Torrieri, {\it {Sound waves and vortices in a
  polarized relativistic fluid}},  {\em Phys. Rev.} {\bf D96} (2017), no.~7
  076016, [\href{http://arxiv.org/abs/1703.03079}{{\tt arXiv:1703.03079}}].

\bibitem{Montenegro:2018bcf}
D.~Montenegro and G.~Torrieri, {\it {Causality and dissipation in relativistic
  polarizeable fluids}},  \href{http://arxiv.org/abs/1807.02796}{{\tt
  arXiv:1807.02796}}.

\bibitem{Florkowski:2017ruc}
W.~Florkowski, B.~Friman, A.~Jaiswal, and E.~Speranza, {\it {Relativistic fluid
  dynamics with spin}},  {\em Phys. Rev.} {\bf C97} (2018), no.~4 041901,
  [\href{http://arxiv.org/abs/1705.00587}{{\tt arXiv:1705.00587}}].

\bibitem{Florkowski:2017dyn}
W.~Florkowski, B.~Friman, A.~Jaiswal, R.~Ryblewski, and E.~Speranza, {\it
  {Spin-dependent distribution functions for relativistic hydrodynamics of
  spin-1/2 particles}},  {\em Phys. Rev.} {\bf D97} (2018), no.~11 116017,
  [\href{http://arxiv.org/abs/1712.07676}{{\tt arXiv:1712.07676}}].

\bibitem{Florkowski:2018myy}
W.~Florkowski, E.~Speranza, and F.~Becattini, {\it {Perfect-fluid hydrodynamics
  with constant acceleration along the stream lines and spin polarization}},
  {\em Acta Phys. Polon.} {\bf B49} (2018) 1409,
  [\href{http://arxiv.org/abs/1803.11098}{{\tt arXiv:1803.11098}}].

\bibitem{Florkowski:2018fap}
W.~Florkowski and R.~Ryblewski, {\it {Hydrodynamics with spin --- pseudo-gauge
  transformations, semi-classical expansion, and Pauli-Lubanski vector}},
  \href{http://arxiv.org/abs/1811.04409}{{\tt arXiv:1811.04409}}.

\bibitem{Florkowski:2018ahw}
W.~Florkowski, A.~Kumar, and R.~Ryblewski, {\it {Thermodynamic versus kinetic
  approach to polarization-vorticity coupling}},  {\em Phys. Rev.} {\bf C98}
  (2018), no.~4 044906, [\href{http://arxiv.org/abs/1806.02616}{{\tt
  arXiv:1806.02616}}].

\bibitem{deGroot:1980}
S.~de~Groot, W.~van Leeuwen, and C.~van Weert, {\it {Relativistic Kinetic
  Theory: Principles and Applications}},  {\em {\it North-Holland, Amsterdam}}
  (1980).

\bibitem{Vasak:1987um}
D.~Vasak, M.~Gyulassy, and H.~T. Elze, {\it {Quantum Transport Theory for
  Abelian Plasmas}},  {\em Annals Phys.} {\bf 173} (1987) 462--492.

\bibitem{Florkowski:1995ei}
W.~Florkowski, J.~Hufner, S.~P. Klevansky, and L.~Neise, {\it {Chirally
  invariant transport equations for quark matter}},  {\em Annals Phys.} {\bf
  245} (1996) 445--463, [\href{http://arxiv.org/abs/hep-ph/9505407}{{\tt
  hep-ph/9505407}}].

\bibitem{Itzykson:1980rh}
C.~Itzykson and J.~B. Zuber, {\em {Quantum Field Theory}}.
\newblock International Series In Pure and Applied Physics. McGraw-Hill, New
  York, 1980.

\end{thebibliography}\endgroup
\bibliographystyle{JHEP}

\end{document}